\documentclass[pre,aps,preprint,showpacs]{revtex4}
\usepackage{amssymb}
\usepackage{latexsym}
\usepackage{amsfonts}
\usepackage{graphicx}
\usepackage{epsfig}
\usepackage{amsmath}
\usepackage{float}
\usepackage{verbatim}

\begin{document}

\title{Strategy of Competition between Two Groups based on a Contrarian
  Opinion Model } 

\author{Qian Li$^1$}\email{liqian@bu.edu}

\author{Lidia A. Braunstein,$^{2,1}$}

\author{Shlomo Havlin,$^3$}

\author{H.~Eugene Stanley$^1$}

\affiliation{$^1$Department of Physics and Center for Polymer Studies,
  Boston University, Boston, MA 02215, USA \\ $^2$Instituto
  de Investigaciones F\'isicas de Mar del Plata (IFIMAR)-Departamento
  de F\'isica, Facultad de Ciencias Exactas y Naturales, Universidad
  Nacional de Mar del Plata-CONICET, Funes 3350, (7600) Mar del Plata,
  Argentina.\\ $^3$Department of Physics, Bar Ilan University, Ramat
  Gan, Israel}

\date{\today}

\begin{abstract}

We introduce a contrarian opinion (CO) model in which a fraction $p$
of contrarians within a group holds a strong opinion opposite to the
opinion held by the rest of the group. At the initial stage, stable
clusters of two opinions, $A$ and $B$ exist. Then we introduce
contrarians which hold a strong $B$ opinion into the opinion $A$
group. Through their interactions, the contrarians are able to
decrease the size of the largest $A$ opinion cluster, and even destroy
it. We see this kind of method in operation, {\it e.g} when companies
send free new products to potential customers in order to convince
them to adopt the product and influence others. We study the CO model,
using two different strategies, on both Erd\"{o}s-R\'{e}nyi and
scale-free networks. In strategy I, the contrarians are positioned at
random. In strategy II, the contrarians are chosen to be the highest
degrees nodes. We find that for both strategies the size of the
largest $A$ cluster decreases to zero as $p$ increases as in a phase
transition. At a critical threshold value $p_c$ the system undergoes a
second-order phase transition that belongs to the same universality
class of mean field percolation. We find that even for an
Erd\"{o}s-R\'{e}nyi type model, where the degrees of the nodes are not
so distinct, strategy II is significantly more effctive in reducing
the size of the largest $A$ opinion cluster and, at very small values
of $p$, the largest $A$ opinion cluster is destroyed.

\end{abstract}

\pacs{89.75.Hc, 89.65.-s, 64.60.-i, 89.75.Da}

\maketitle

\section{Introduction}

Competition between two groups or among a larger number of groups is
ubiquitous in business and politics: the decades-long battle between
the Mac and the PC in the computer industry, between Procter \& Gamble
and Unilever in the personal products industry, among all major
international and local banks in the financial market, and among
politicians and interest groups in the world of governance. All
competitors want to increase the number of their supporters and thus
increase their chances of success. In gathering supporters,
competitors put much effort into persuading skeptics and those
opponents who may actually be potential supporters. This kind of
activity is normally modeled as a dynamic process on a complex network
in which the nodes are the agents and the links are the interactions
between agents. The goal of these models is to understand how an
initially disordered configuration can become an ordered configuration
through the interaction between agents. In the context of social
science, order means agreement and disorder means disagreement
\cite{socialCastellano, socialGalam}. Most of these models---e.g., the
Sznajd model \cite{Sznajd}, the voter model \cite{voter1, voter2}, the
majority rule model \cite{majority1, majority2}, and the social impact model
\cite{social1, social2}---are based on two-state spin systems which
tend to reduce the variability of the initial state and lead to a
consensus state in which all the agents share the same
opinion. However this consensus state is not very realistic, since in
many real competitions there are always at least two groups that
co-exist at the same time.

Recently a non-consensus opinion (NCO) model \cite{NCO} was developed,
where two opinions $A$ and $B$ compete and reach a non-consensus
stable state. At each time step each node adopts the opinion of the
majority in its ``neighborhood,'' which consists of its nearest
neighbors and itself. When there is a tie, the node does not change
its state. Considering also the node's own opinion leads to the
non-consensus state. The dynamics are such that a steady state in
which opinions $A$ and $B$ coexist is quickly reached. It was
conjectured, and supported by intensive simulations \cite{NCO}, that
the NCO model in complex networks belongs to the same universality
class as percolation \cite{NCO, percolation, quenched_shlomo}.

Here we test how competition strategies are affected when
``contrarians'' are introduced. Contrarians are agents who hold a
strong opinion that is opposition to an opinion held by the rest of
the group, and are not influenced by their opinion but can influence
them. We develop a spin-type contrarian opinion (CO) model in which
contrarian agents are introduced into the steady state of the NCO
model. The goal of the contrarians is to change the opinions of the
current supporters of the rival group \cite{contrarian}. We see this
strategy in operation, for example, when companies send free new
products to potential customers in order to convince them to adopt the
product and encourage their friends to do the same. We can observe it
also in political campaigns when candidates ``bribe'' voters by
offering favors. The questions we ask in our model are as follows. Do
these free products and bribes work and how? Who are the best
individuals to chose as contrarians in order to make most impact on
opinion change.

In this paper we introduce, into group $A$, a fraction $p$ of
contrarians, which are defined to be agents that hold a strong $B$
opinion, who will influence those who hold the $A$ opinion to change
their opinion to $B$. We study two different strategies of introducing
contrarians, (I) random choosing of contrarians, and (II) targeted. We
study these strategies on two types of networks, Erd\"{o}s-R\'{e}nyi
(ER) \cite{ER1} and scale free (SF) \cite{SF}. We find, for both
strategies, that the relative size of the largest cluster in state $A$
undergoes a second order phase transition at a critical fraction of
contrarians $p_c$. Moreover we find that, for both networks analyzed
here, the targeted strategy is much more efficient than the random
strategy. Our results indicate that the observed second order phase
transition can be mapped into mean field percolation.

\section{The CO Model}

In the NCO model\cite{NCO}, initially, a fraction $f$ of agents with
$A$ opinions and $1-f$ with $B$ opinions are selected at random. At
each time step, each network node adopts the majority opinion based on the opinions of its
neighbors and itself. All updates are performed simultaneously at each
time step until a steady state is reached in which both opinions
coexist, which occurs for $f$ above a critical threshold $f\equiv
f_c$.

\begin{figure}[ht]
\includegraphics[width=6.5cm,height=4.5cm]{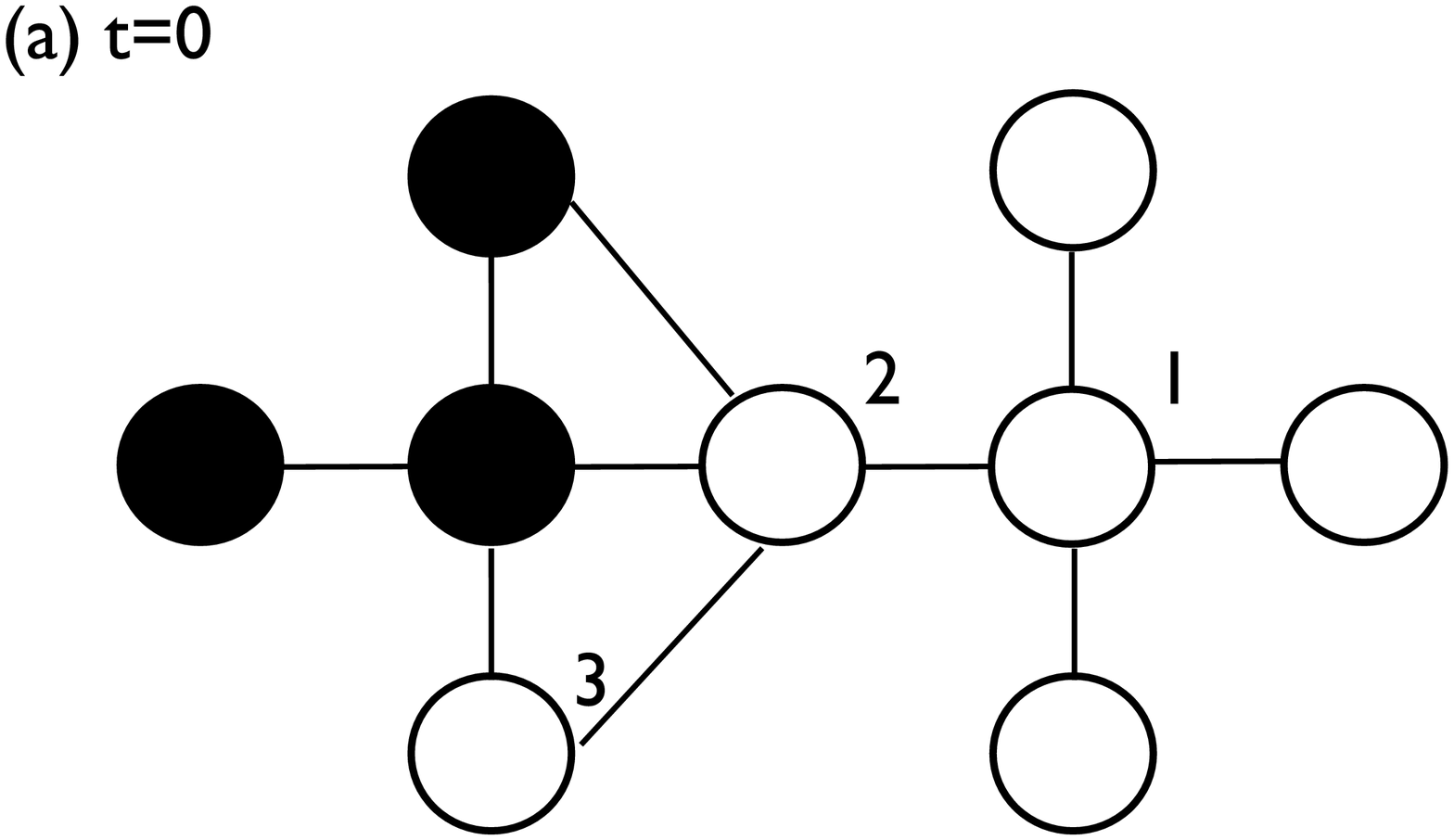}
\hspace{1.5cm}
\includegraphics[width=6.5cm,height=4.5cm]{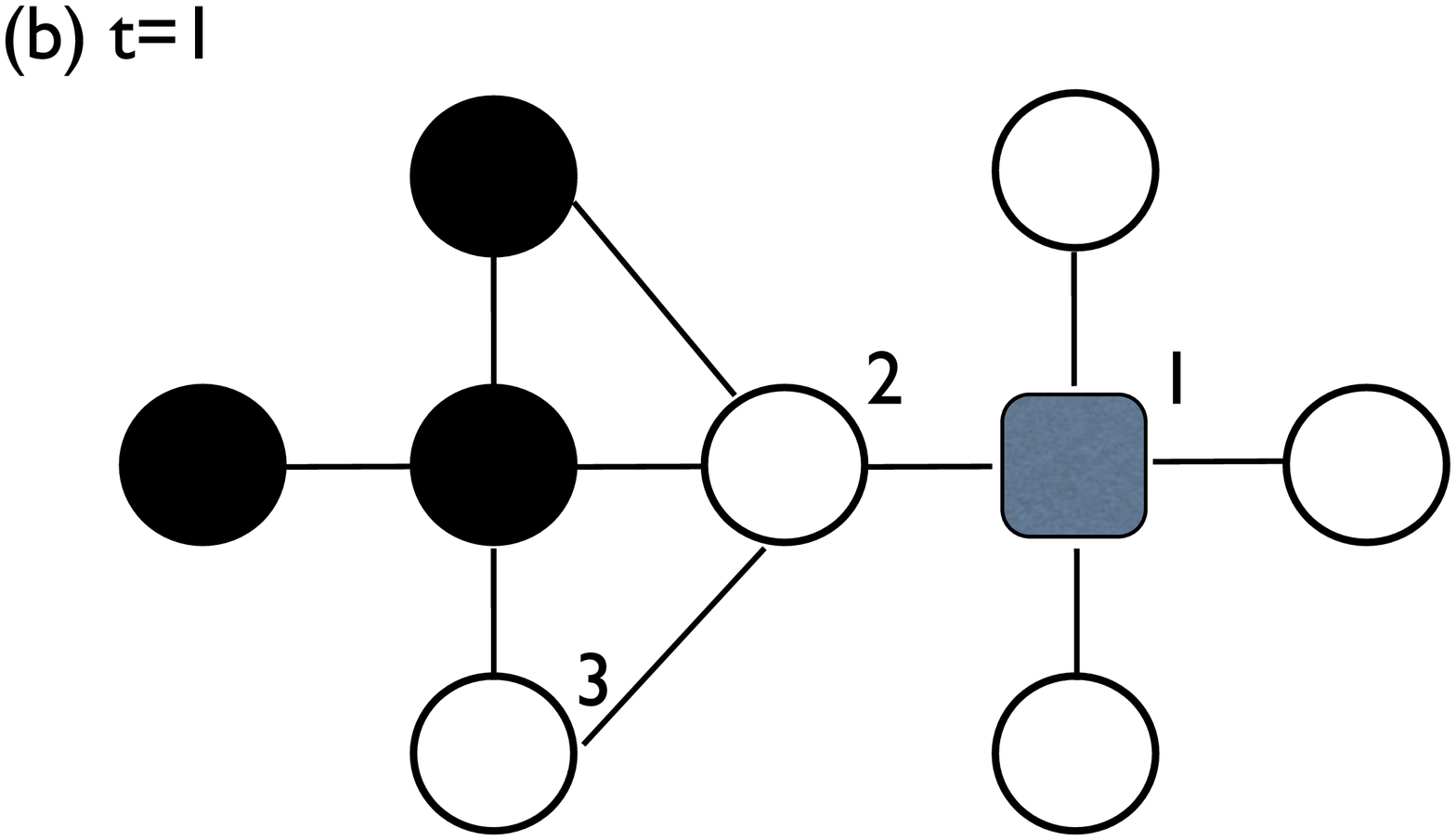}\\
\vspace{2.0cm}

\includegraphics[width=6.5cm,height=4.5cm]{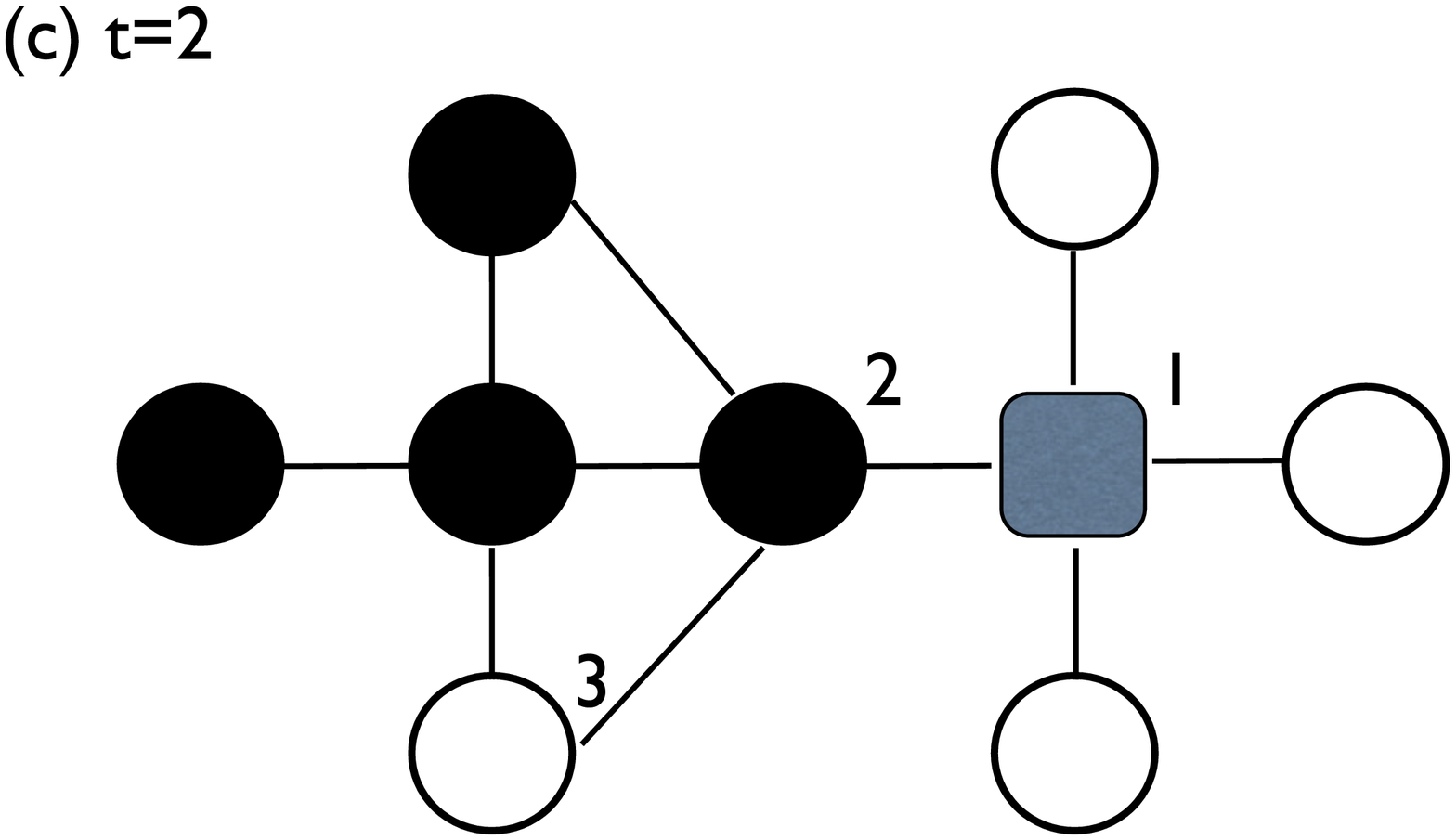}
\hspace{1.5cm}
\includegraphics[width=6.5cm,height=4.5cm]{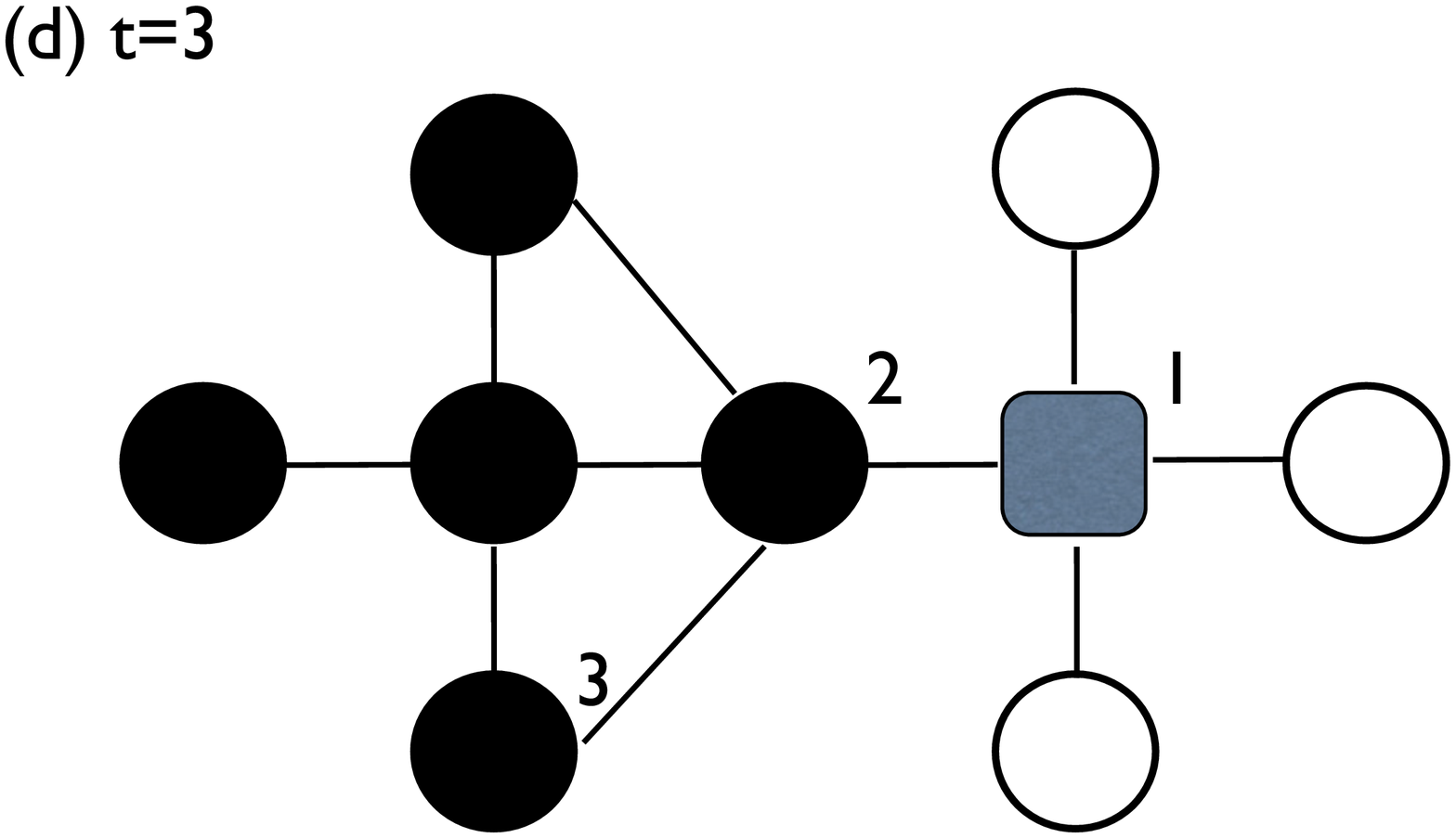}
\vspace{0.5cm}

\caption{Schematic plot of the dynamics of the CO model showing the
  approach to a stable state on a network with $N=9$ nodes. (a) At
  $t=0$, we have a stable state where opinion $A$ (open circle) and
  opinion $B$ (filled circle) coexist. (b) At $t=1$, we change node $1$
  into a contrarian (filled square), which will hold $B$ opinion. Node
  $2$ is now in the local minority opinion while the remaining nodes are
  not.  Notice that node $1$ is a contrarian and even if he is in the
  local minority he does not change his opinion.  At the end of this
  simulation step, node $2$ is converted into $B$ opinion. (c) At $t=2$,
  node $3$ is in the local minority opinion and therefore will be
  converted into $B$ opinion. (d) At $t=3$, the system reaches a stable
  state where the system breaks into four disconnected clusters, one of
  them composed of six $B$ nodes and the other three with one $A$
  node. \label{fig:f.12}}
\end{figure}

In our CO model, the initial state is the final state of the NCO
model. Above $f_c$ we have stable clusters of both $A$ or $B$ opinions
in a network of $N$ agents. We choose, initially, a fraction $p$ of
$A$ opinion agents that are changed into $B$ opinion agents and so
become contrarians. By contrarian we mean that, they will never
change their opinion but they can influence others.  Then we use again the
NCO dynamics to reach a new steady state. In the new steady state the
agents form again clusters of two opposite opinions above a certain
threshold $f_c$ that now depends on $p$. Because of the contrarians
of type $B$, the $A$ clusters become smaller and the $B$ clusters
increase. In Fig.~\ref{fig:f.12} we show a schematic plot of the
dynamics of the CO model.

We use both random and targeted strategies to choose a fraction $p$
of $A$ agents that flip into state $B$, and we analyze the results on
ER and SF networks. In strategy I we randomly choose a fraction $p$
of A agents, and in strategy II we choose the top $p$ percent of the
highest degree $A$ agents, to become contrarians. Notice that the
contrarians act as a quenched disorder in the network
\cite{quenched_barabasi, quenched_shlomo}.

\section{Simulation Results}

We perform simulations of the CO model in complex networks, on both ER
and SF networks. ER networks are characterized by a Poisson degree
distribution with $P(k)=e^{- \langle k \rangle} \langle k \rangle^k
/k!$, where $k$ is the degree of a node and $\langle k \rangle$ is the
average degree. In SF networks the degree distribution is given by
$P(k)\sim k^{-\lambda}$, for $k_{\rm min}\leq k \leq k_{\rm max}$,
where $k_{\rm min}$ is the smallest degree, $k_{\rm max}$ is the
degree cutoff and $\lambda$ is the exponent characterizing the
broadness of the distribution. In all our simulations we use the
natural cutoff, which is controlled by $N^{1/(\lambda-1)}$
\cite{percolation_shlomo}. We performed all the simulations for $10^3$
network realizations.

\subsection{CO model on ER networks}

\begin{figure}[ht]
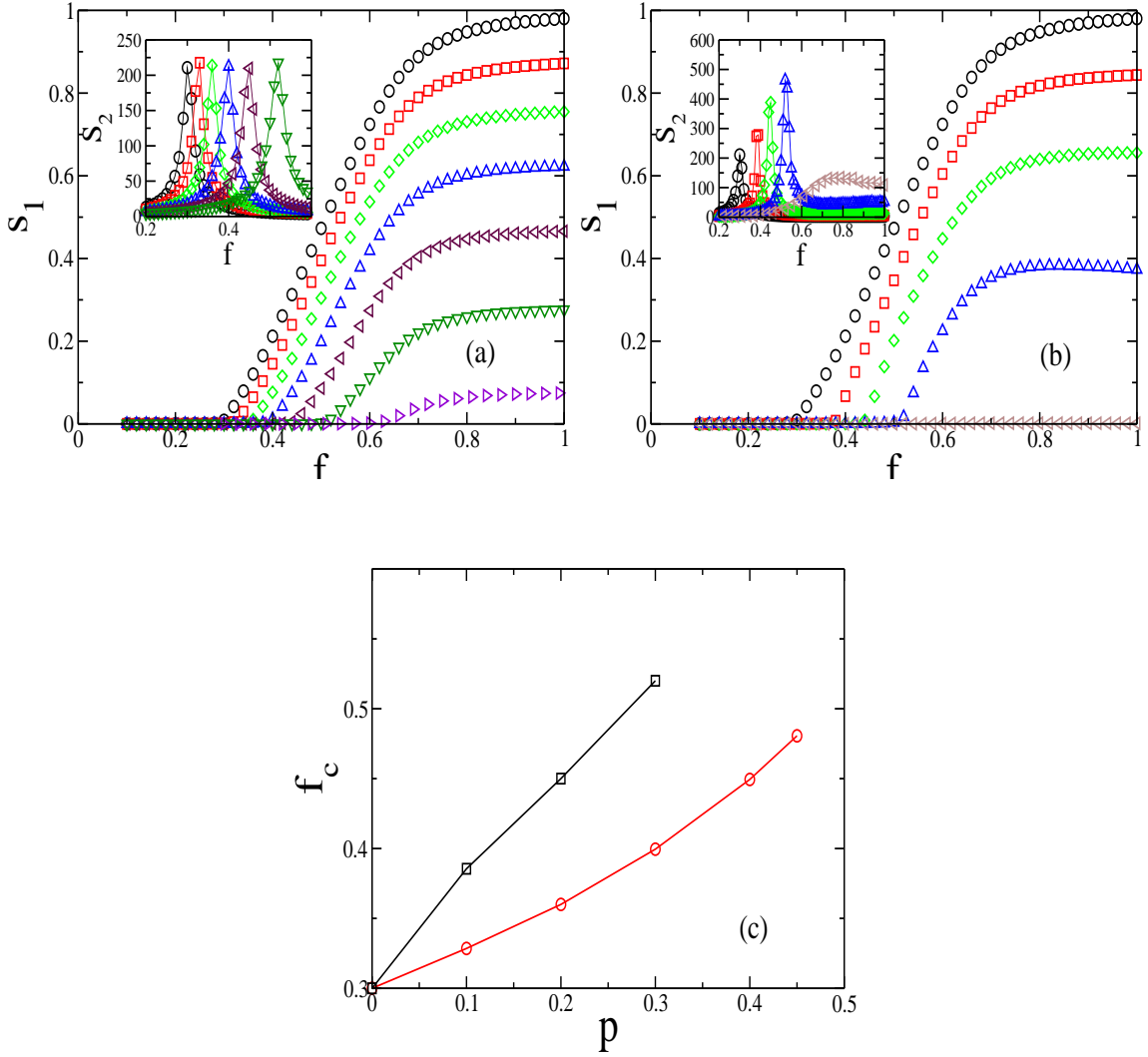

\includegraphics[width=7.5cm,height=6.5cm, angle=0]{fig1_a.eps}
\includegraphics[width=7.5cm,height=6.5cm, angle=0]{fig1_b.eps}\\
\vspace{1cm}
\includegraphics[width=7.5cm,height=6.5cm, angle=0]{fig1_c.eps}

\caption{Plot of $s_1$ as a function of $f$ for different values of
  $p$ for ER networks with $\langle k \rangle=4$, and $N=10^5$.  (a)
  Strategy I, $p =0$ ($\circ$), $0.1$ ($\Box$) $0.2$ ($\diamond$),
  $0.3$ ($\bigtriangleup$), $0.4$ ($\triangleleft$), $0.5$
  ($\triangledown$) and $p=p^*=0.6$ ($\triangleright$). (b) Strategy
  II, $p =0$ ($\circ$), $0.1$ ($\Box$), $0.2$ ($\diamond$), $0.3$
  ($\bigtriangleup$) and $p=p^*=0.4$ ($\triangleleft$). In the inset
  we plot, using the same symbols as in the main figure, $S_2$ as a
  function of $f$ for both strategies. (c) Plot of $f_c$ as a function
  of $p$ for strategy I ($\circ$) and II ($\Box$).\label{fig:f.2}}
\end{figure}

\begin{figure}[ht]
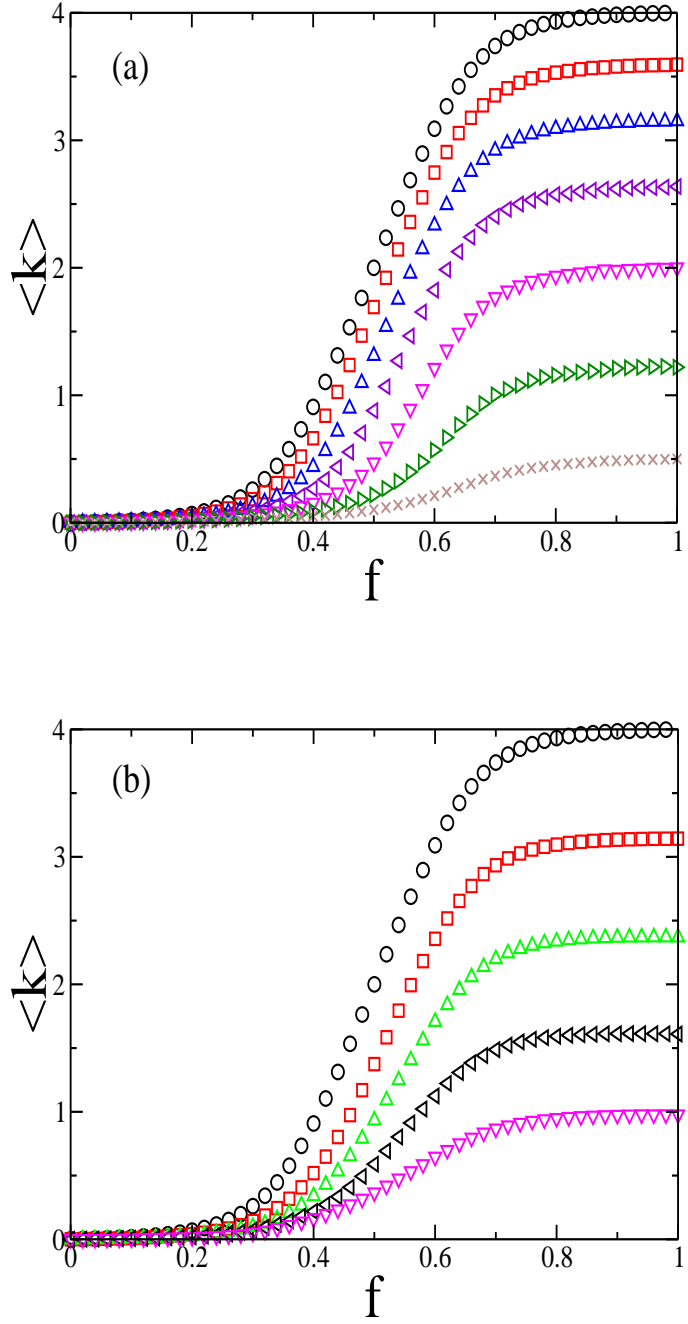

\includegraphics[width=9cm,height=8cm]{fig_er_ra_con.eps}\\
\vspace{1.5cm}
\includegraphics[width=9cm,height=8cm]{fig_er_ta_con.eps}
\caption{Plot of $\langle k\rangle$ as a function of $f$ for different
  value of $p$ for ER networks with $\langle k \rangle=4$,
  $N=10^5$. (a) strategy I, $p =0$ ($\circ$), $0.1$ ($\Box$) $0.2$
  ($\bigtriangleup$), $0.3$ ($\triangleleft$), $0.4$
  ($\triangledown$), $0.5$ ($\triangleright$) and $0.6$ (x) and (b)
  strategy II,  $p =0$ ($\circ$), $0.1$ ($\Box$) $0.2$
  ($\bigtriangleup$), $0.3$ ($\triangleleft$) and $0.4$
  ($\triangledown$).\label{fig:f.CON}}
\end{figure}

We denote by $S_1$ the size of the largest $A$ cluster in the steady
state. We study the effect of the contrarians. In Fig.~\ref{fig:f.2}
we plot $s_1\equiv S_1/N$ as a function of $f$ for different values of
$p$ for ER networks under both random and targeted strategies. The
plot shows that there exists a critical value $f\equiv f_c$ that
depends on $p$, below which $s_1$ approaches zero.  As $p$ increases,
the largest cluster becomes significantly smaller as well as less
robust, as can been seen from the shift of $f_c$ to the larger value.
In the inset of Fig.~\ref{fig:f.2}, we plot the size of the second
largest $A$ cluster, $S_2$, as a function of $f$ for different values
of $p$. $S_2$ shows a sharp peak characteristic of a second order
phase transition, where $s_1$ is the order parameter and $f$ is the
control parameter. Above a certain value of $p \equiv p^*$, the phase
transition does not occur because, above $p^*$, the average
connectivity of the $A$ nodes decreases dramatically, the networks
break into small clusters and no giant component of opinion $A$
appears. In Fig.~\ref{fig:f.CON} we show, for both strategies, the
average degree $\langle k \rangle$ of the $A$ opinion agents as a
function of $f$ for different values of $p$. As shown in \cite{NCO}
for $p=0$, $\langle k\rangle$ has a significant increase above $f=0.5$
where the nodes with opinion $A$ are majority. This is because when
these nodes are in the minority group, they have a small average
connectivity since the minority group doesn't include high degree
nodes. This process is analogous to targeted removing of the high
degree nodes. Only when they become majority nodes, close to $f=1$,
the original connectivity of the full network is recovered. However,
as $p$ increases, we never reach the original average degree of the
full network because increasing $p$ is equivalent to increasing the
quenched disorder. It is known that for ER networks the transition is
lost when $\langle k\rangle < 1$ \cite{ER1}. As we can see from the
plots, for $p^* \approx 0.6$ (strategy I) and $p^*\approx 0.4$
(strategy II), $\langle k\rangle$ drops below $1$, and then the giant
component cannot be sustained.

The loss of robustness is significantly more pronounced in the
targeted strategy compared to the random strategy, as seen in
Fig.~\ref{fig:f.2}(c), where we plot $f_c$ as a function of $p$
for both strategies. From this plot we can see that the targeted
strategy is significantly more efficient to annihilate the opinion $A$
clusters than the random strategy. For example, for $p=0.2$, the
network is 25\% less robust in the targeted strategy compared to the
random one. The reason is that the initial state of our model is the
final state of the NCO model, which above its threshold has clusters
of nodes $A$ of all sizes. Thus under the random strategy we select
nodes at random that are mainly in small $A$ clusters. Under the
targeted strategy the selection of contrarians from the nodes of
highest degree places them mainly in the largest initial $A$ cluster
where they can influence more than if they were isolated in smaller
clusters, as in the random strategy. The high degree nodes shorten the
distances between all the pairs of nodes in a cluster, which allows
them to interact more easily. Also, because they have many neighbors,
they can influence the opinions of other $A$ nodes more efficiently.

\begin{figure}[ht]
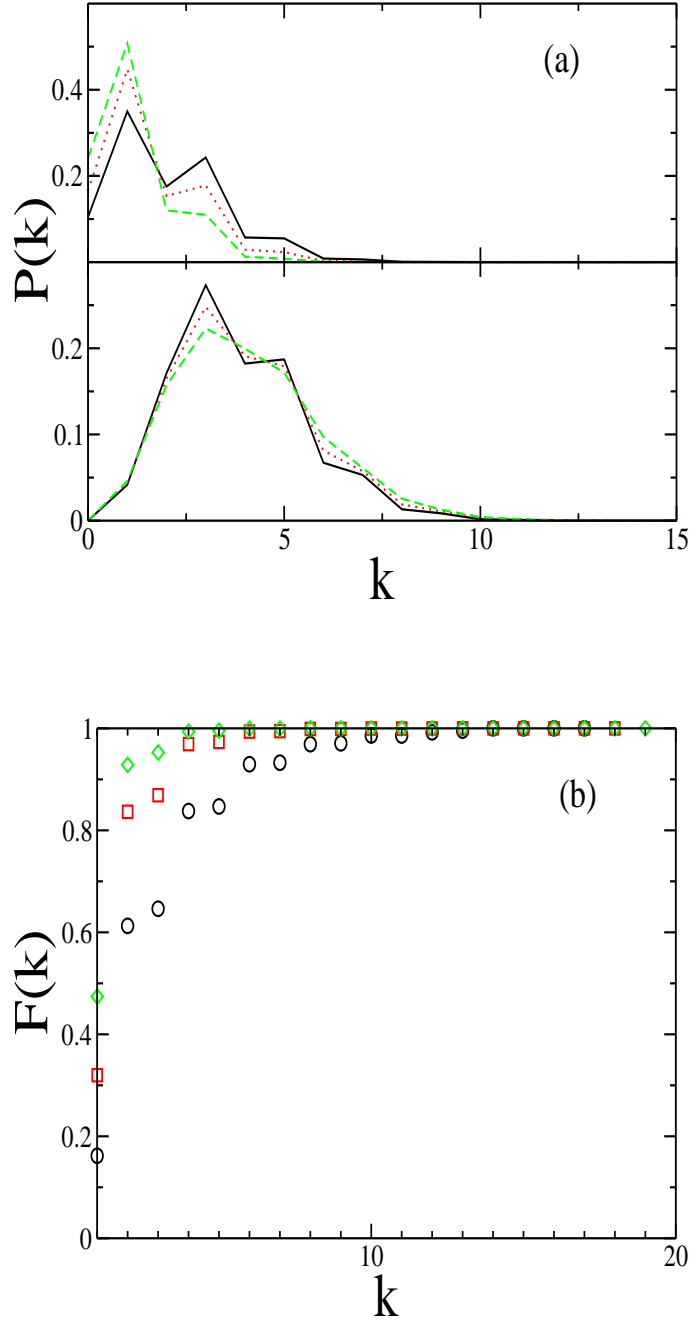

\includegraphics[width=9cm,height=8cm, angle=0]{Pk.eps}\\
\vspace{1.5cm}
\includegraphics[width=9cm,height=8cm]{Fk_ER4.eps}
\caption{(Color online) For ER network with $\langle k \rangle=4$ and
  $N=10^5$, (a) Degree Distribution $P(k)$ of $A$ nodes as a function
  of $k$ in our initial configuration with different values of $f$,
  $f=0.35$ (full line), $f=0.4$ (dotted line), and $f=0.45$ (dashed
  line). In the top, we show $P(k)$ as a function of $k$ of the finite
  $A$ clusters, and in the bottom the same for the largest $A$
  cluster.  (b) Plot of $F(k)$ as a function of $k$ for different
  value of $f$, $f =0.35$ ($\circ$), $0.4$ ($\Box$) and $0.45$
  ($\diamond$). \label{fig:f.FER}}
\end{figure}

In order to verify the above arguments, we compute, for our initial condition
($p=0$) before adding the contrarians, the degree distribution of
nodes $A$ inside and outside the largest cluster.  In
Fig.~\ref{fig:f.FER}(a) we plot the degree distributions $P(k)$ of $A$
nodes inside and outside the largest cluster for three different
values of $f$ above the threshold of the NCO model. Notice that the
nodes outside the largest cluster have a degree distribution with a
high probability of low connectivity.  The probability of low
connectivity increases as $f$ increases, and thus under a targeted
strategy the nodes in those small clusters are almost never designated as
contrarians. Thus nodes in the largest component are more likely to be
selected under a targeted strategy than under a random one. 

In order to further test our assumption, we compute the fraction
$F(k)$, defined as the ratio of the number of nodes with degree $k$ in
the largest $A$ cluster and the total number of nodes in all the $A$
clusters with the same degree.  When $F(k)\to 1$, all the nodes with
degree $k$ are in the largest $A$ cluster.  In
Fig.~\ref{fig:f.FER}(b), we plot $F(k)$ as a function of $k$ for
different values of $f$.  As $k$ increases, we see that $F(k)\to 1$ is
faster for increasing $f$ because the larger $f$ is, the larger $S_1$
will be.  Thus the highest degree nodes belong to the largest cluster
and the lower degree nodes are less likely to be in the largest
cluster. This explains why a targeted strategy is significantly more
efficient than a random one.

\begin{figure}[ht]
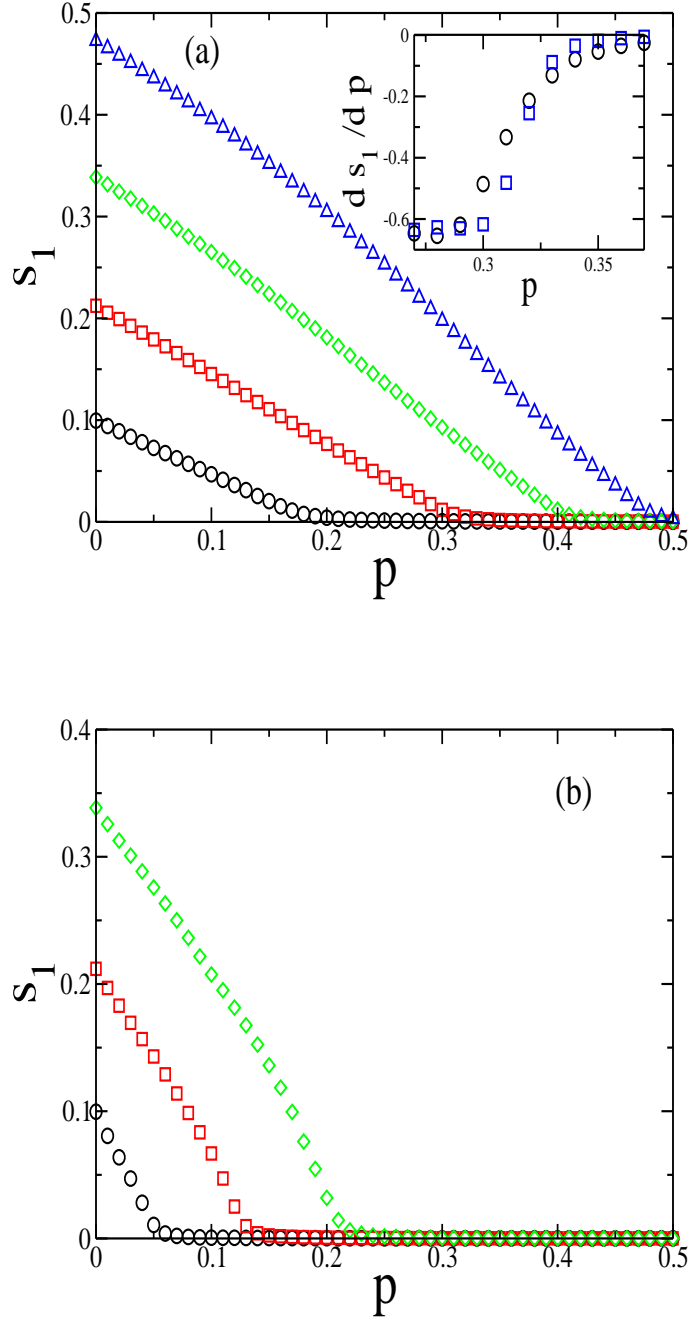

\includegraphics[width=9cm,height=8cm]{fig2_a.eps}\\
\vspace{1.5cm}
\includegraphics[width=9cm,height=8cm]{fig2_b.eps}
\caption{(Color online) Plot of $s_1$ as a function of $p$ for
  different values of $f$ for ER network with $\langle k \rangle=4$
  and $N=10^5$. (a) Strategy I, $f =0.35$ ($\circ$), $0.4$ ($\Box$)
  $0.45$ ($\diamond$) and $0.5$ ($\bigtriangleup$). (b) Strategy II
  for $f =0.35$ ($\circ$), $0.4$ ($\Box$) and $0.45$ ($\diamond$).  In
  the inset of (a), we plot the first derivative of $s_1=S_1/N$ with
  respect to $p$ ($d s_1 /d p$) with different system size, $N=10^5$
  ($\circ$) and $N=10^6$ ($\Box$).\label{fig:f.4}}
\end{figure}

Because $p$ is our main parameter, we next investigate the behavior
of the system as a function of $p$ for different values of $f$.  In
Fig.~\ref{fig:f.4} we plot $s_1$ as a function of $p$ for fixed $f$
for ER networks under both strategies.  From the plot we can see that
$S_1$ is more robust as $f$ is larger, and the behavior of the curve is
again characteristic of a second order phase transition.  However this curve
seems to be smoother than the transition found above (see
Fig.~\ref{fig:f.2}) with $f$ as the control parameter.

\begin{figure}[ht]
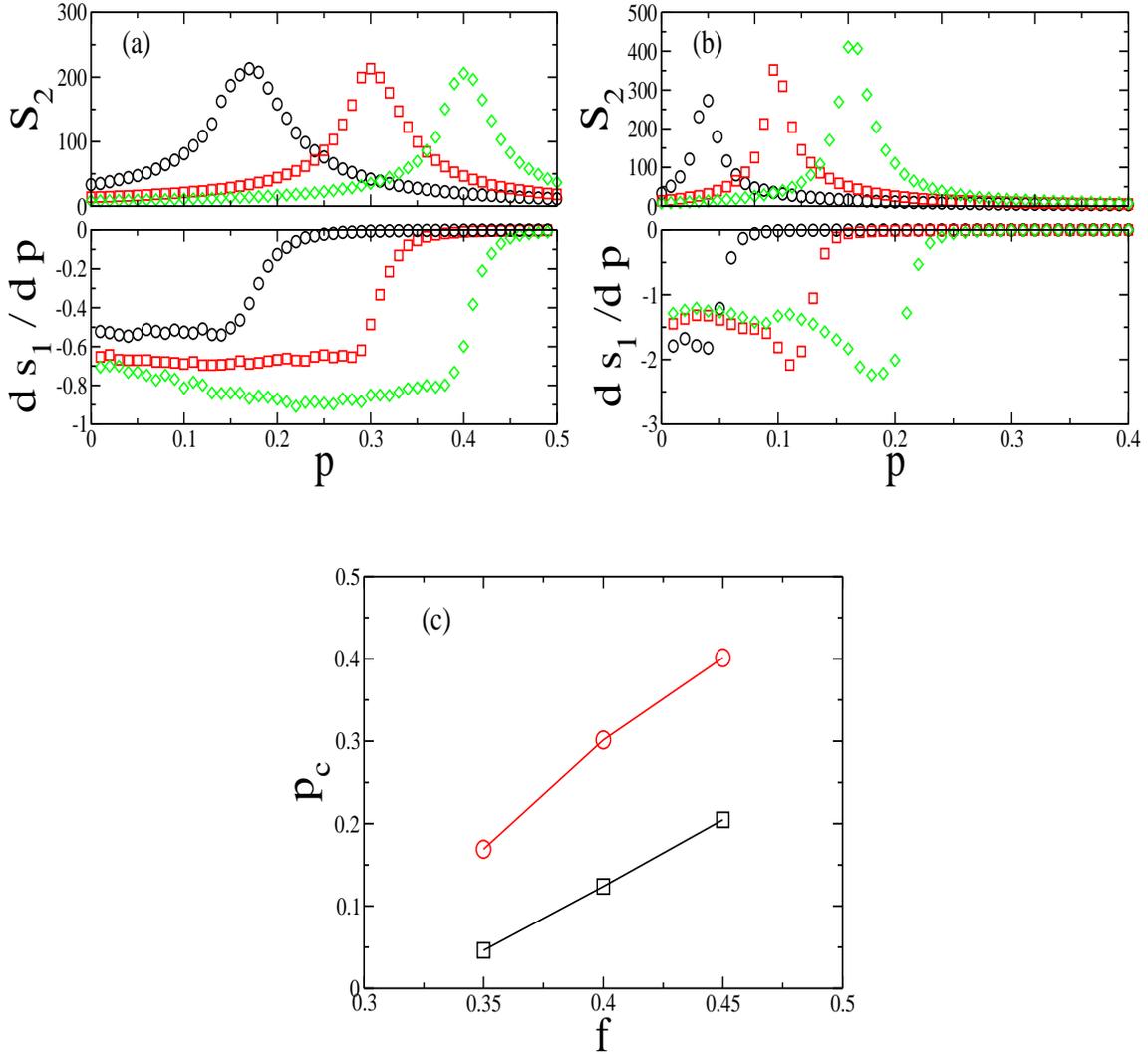

\vspace{1.5cm}
\includegraphics[width=7.5cm,height=6.5cm, angle=0]{fig3_a.eps}
\includegraphics[width=7.5cm,height=6.5cm, angle=0]{fig3_b.eps}\\
\vspace{1cm}
\includegraphics[width=7.5cm,height=6.5cm, angle=0]{fig3_c.eps}
\caption{(Color online) Plot of $S_2$ as a function of $p$ (top) and
  $d s_1 /d p$ as a function of $p$ (bottom) for different values of
  $f$ for (a) strategy I and (b) strategy II, $f =0.35$ ($\circ$),
  $0.4$ ($\Box$) $0.45$ ($\diamond$) for ER networks with $\langle k
  \rangle=4$, $N=10^5$. We can see that in both cases the peak of
  $S_2$ coincides with the position of the jump, also indicating a
  second order phase transition. In (c) we plot $p_c$ as function of
  $f$ for both strategies, strategy I ($\circ$) and strategy II
  ($\Box$).\label{fig:f.6}}
\end{figure}

In the inset of Fig.~\ref{fig:f.4} (a) we plot the first derivative of
$s_1$ with respect to $p$ for two different system sizes for
$f=0.4$.  We can see a jump that becomes sharper as the system size
increases. We find the same behavior for other values of $f$ above the
threshold. In Figs.~\ref{fig:f.6} (a) and \ref{fig:f.6} (b) we show
$S_2$ and the first derivative of $s_1$ with respect to $p$ for
$N=10^5$ and for different values of $f$. We find that the peak of
$S_2$ and the jump of the derivative of $s_1$ occurs at the same value
of $p$. This behavior is characteristic of a second order phase
transition, where the peak of $S_2$ indicates the position of the
threshold $p_c$. In Fig.~\ref{fig:f.6}(c) we plot the critical
threshold $p_c$ as a function of $f$ for both
strategies. Comparing the two strategies for the same value of $f$,
strategy II always has the smaller $p_c$. This demonstrates
again that strategy II is better because a very small fraction of
$p$ is enough to destroy the $A$ opinion clusters.

\begin{figure}[ht]
\includegraphics[width=14cm,height=12cm, angle=0]{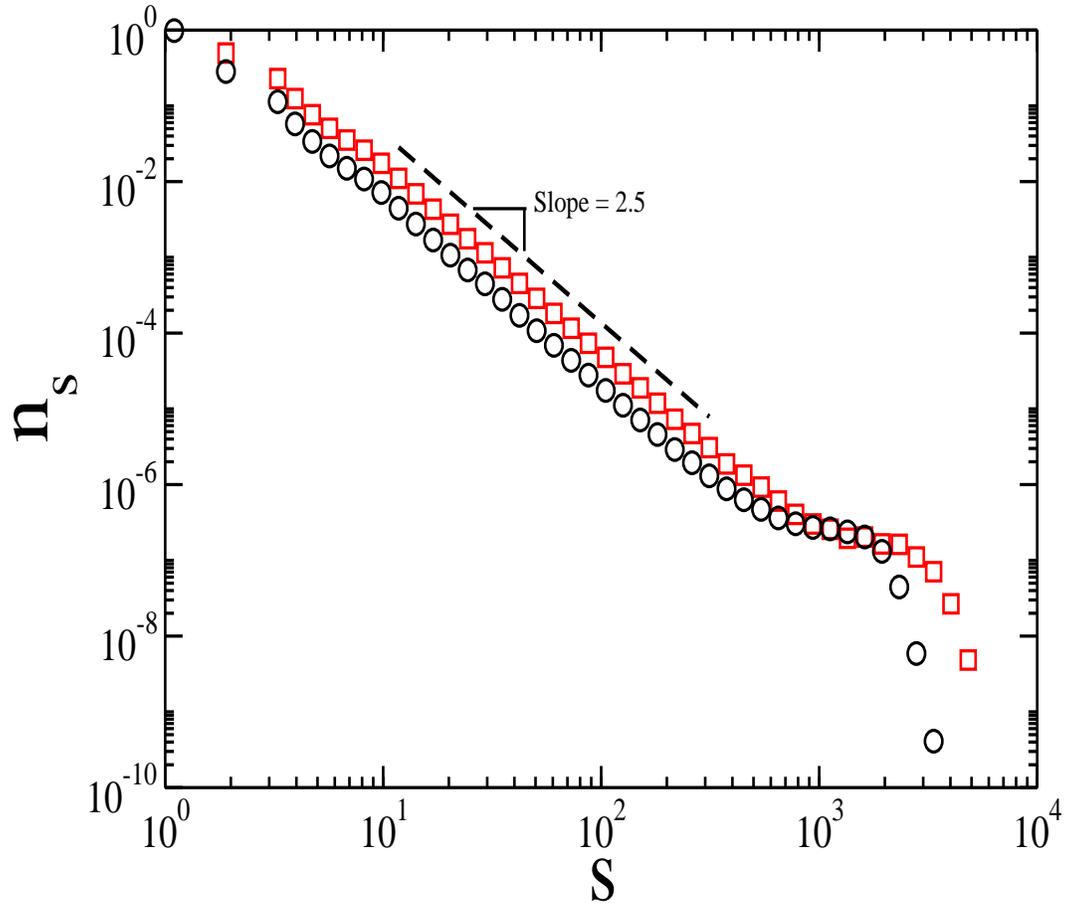}
\caption{(Color online) Plot of $n_s$ as a function of $s$ under
  strategy I ($\circ$) and II ($\Box$) for ER networks with $\langle k
  \rangle=4$ and $N=10^5$ at criticality $p=p_c$. The dashed line
  represents a slope $\tau=2.5$. This simulation were done over $10^5$
  realizations.\label{fig:f.8}}
\end{figure}

Next, we present results indicating that the CO model is in the same
universality class as regular percolation.  Percolation in random
networks ({\it e.g}, ER)\cite{percolation,quenched_shlomo,
  Complexnetwork} predicts that at criticality the cluster size
distribution of finite clusters $n_s \sim s^{-\tau}$ with $\tau=2.5$.
In Fig.~\ref{fig:f.8} we plot $n_s$ for both random and targeted
strategies as a function of $s$ for finite $A$ clusters at
criticality. As we can see for both strategies, $\tau \approx
2.5$. Moreover, from $S_2$ we calculate the exponent $\gamma$, which
represents how the mean finite size diverges with distance to
criticality (not shown here), from which we obtain $\gamma \approx 1$,
as in mean field percolation.  The values of the two exponents we
obtain strongly indicate that our CO model in ER networks is in the
same universality class as mean field percolation in complex networks
below $p^*$ \cite{note}.

\subsection{CO model on SF networks}

Many real social networks are not ER, but instead possess a SF degree
distribution. It is well known that dynamic processes in SF networks
propagate significantly more efficiently
\cite{SFbetter1,SFbetter2,SFbetter3,SFbetter4,SFbetter5, SFbetter6}
than in ER networks. For SF networks we find that the system also
undergoes a second order phase transition as in ER networks with mean
field exponents (not shown here).

\begin{figure}[ht]
\includegraphics[width=10cm,height=9cm, angle=0]{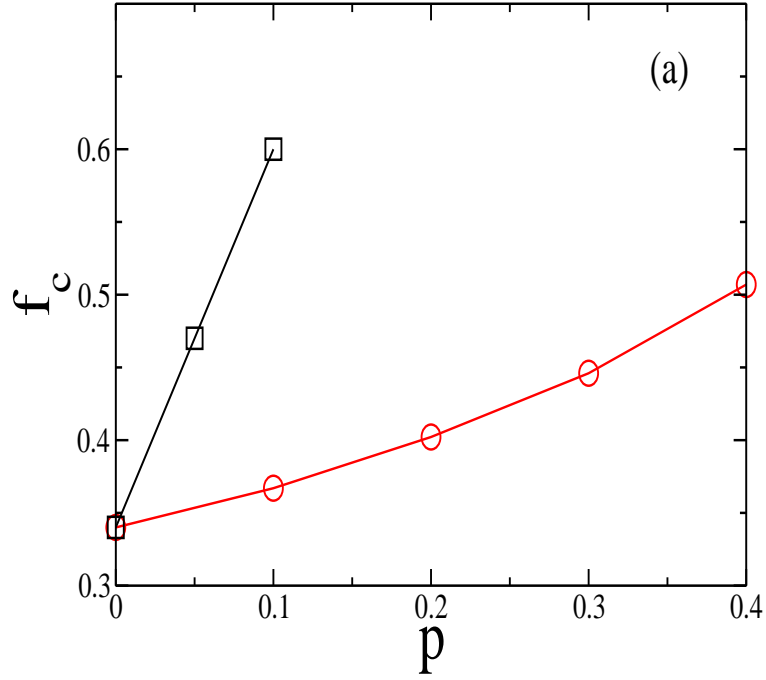}\\
\vspace{1.5cm}
\includegraphics[width=10cm,height=9cm, angle=0]{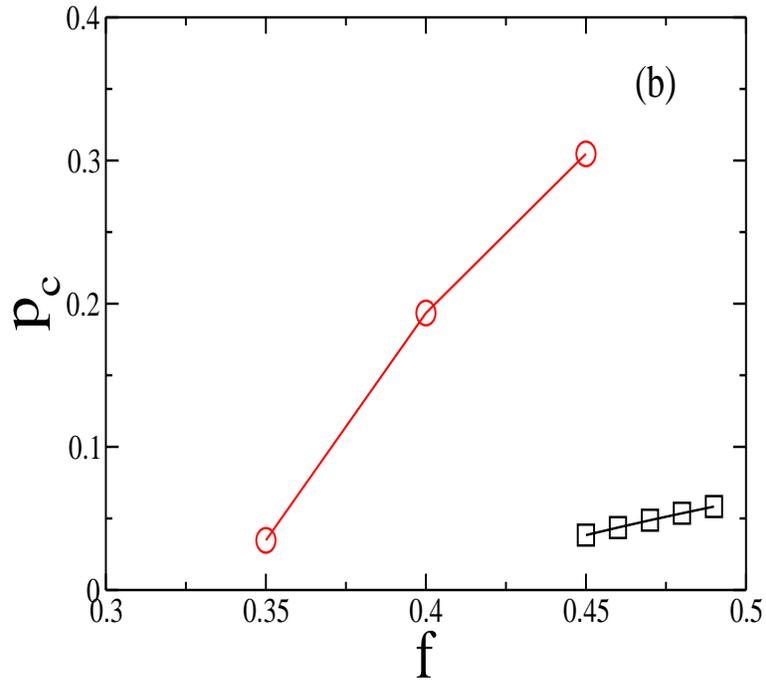}
\caption{(Color online) Plot of (a) $f_c$ as a function of $p$ and (b)
  $p_c$ as a function of $f$ for strategy I ($\circ$) and II ($\Box$)
  for SF networks with $\lambda=3.5$, $k_{\rm min}=2$ and
  $N=10^5$.\label{fig:f.3}}
\end{figure}

\begin{figure}[ht]
\includegraphics[width=14cm,height=12cm]{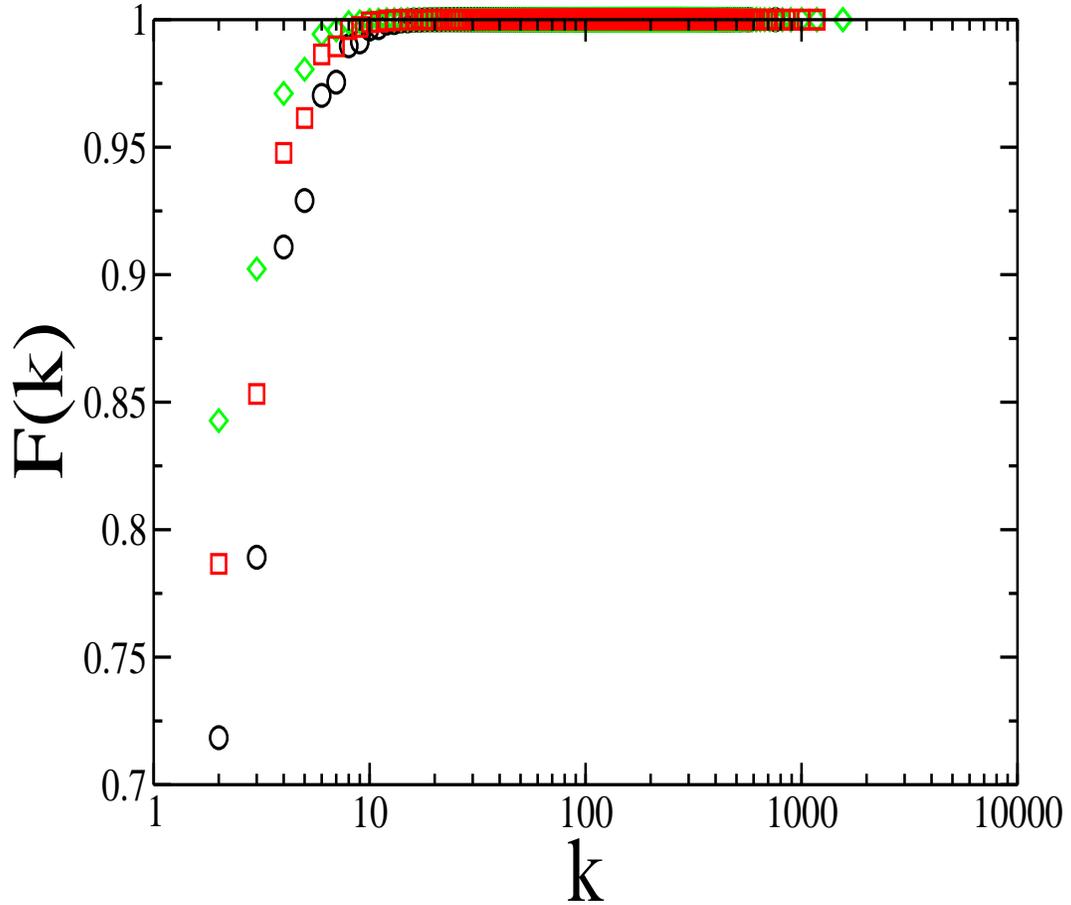}
\caption{(Color online) Plot of $F(k)$ as a function, in linear-log
  scale, of $k$ for ER network with $\langle k \rangle=4$ and $N=10^5$
  with different value of $f$, $f =0.35$ ($\circ$), $0.4$ ($\Box$)
  $0.45$ ($\diamond$). The reason of using linear-log scale is that
  for SF networks $F(k)$ increase very fast for small value of
  $k$ \label{fig:f.10}}
\end{figure}

In Fig.~\ref{fig:f.3} (a) we plot $f_c$ as a function of $p$ for SF
networks with $\lambda =3.5$. For a certain value of $p$, when
$f<f_c$, we lose the largest $A$ cluster. Thus the larger the value of
$f_c$, the less robust the networks. From the plot, we find that for
all values of $p$, strategy II has much larger $f_c$ than strategy
I. This shows that SF networks are significantly less robust under
strategy II than under strategy I, which shows that for SF networks,
strategy II is significantly more efficient compared to strategy I. To
further test our conclusion, in Fig.~\ref{fig:f.3} (b), we plot $p_c$
as a function of $f$ for the same SF networks. As $p_c$ is the minimum
concentration of contrarians needed to destroy the largest $A$
cluster, thus for the same initial condition, the networks are less
robust with smaller $p_c$ than larger one. As shown in
Fig.~\ref{fig:f.3}(b), for the same value of $f$, $p_c$ under strategy
II is always significantly smaller than under strategy I. This result
again support our former conclusion that for SF networks, strategy II
is more efficient than strategy I. As mentioned above, this is because
the targeted strategy sends most of the contrarians into the largest
$A$ cluster. In order to test that, in Fig.~\ref{fig:f.10} we plot
$F(k)$ as a function of $k$ for SF networks. As we can see from
Fig.~\ref{fig:f.10}, almost all of the high degree nodes ($k \gtrsim
10$) belong to the largest $A$ cluster. This behavior is more
pronounced as $f$ increases because $S_1$ increases with $f$.

\subsection{Comparison between ER and SF networks}

If we compare all our results between ER and SF networks, we find that
both strategies are more efficient for SF networks. For example, when we
compare Fig.~\ref{fig:f.10} (a) with Fig.~\ref{fig:f.2} (c) we see that
for all the values of $p$, $f_c$ is larger for SF networks than for ER
networks for both strategies. The main difference between ER and SF
networks is that SF networks possess larger hubs than ER networks, and
thus it is more efficient to destroy the largest $A$ cluster. We also
find that the targeted strategy is more efficient in SF than in ER
networks due to the presence of these large hubs.  For example, when
$p=0.1$, the SF network is 64\% less robust under the targeted
strategy than under the random strategy. In ER networks, for the same
value of $p$, the robustness of the networks decreases only by
17\%. If we compare Fig.~\ref{fig:f.FER} (b) with Fig.~\ref{fig:f.10}
we see that higher degree nodes are more likely to belong to the
largest cluster in SF networks than in ER networks, since $F(k) \to 1$
faster in SF networks compared to ER networks.

\subsection{ Minority vs Majority}
\begin{figure}[ht]
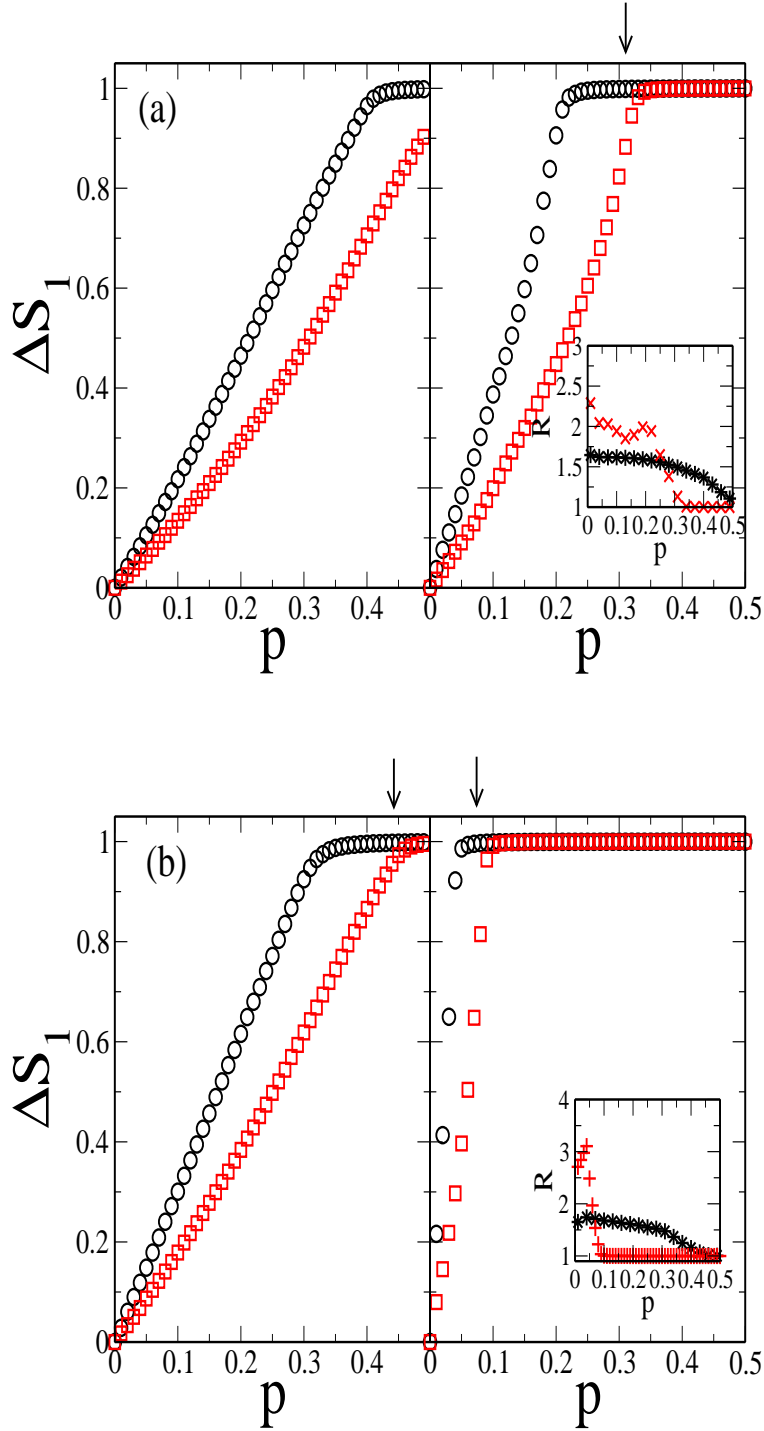

\includegraphics[width=10cm,height=9cm]{fig_er_delts1.eps}\\
\vspace{1cm}
\includegraphics[width=10cm,height=9cm]{fig_sf_delts1.eps}

\caption{(Color online) Plot of $\Delta S_1$ as a function of $p$ for
  $f=0.45$ ($\circ$) and for $f=0.55$ ($\Box$) under strategy I (left
  plot) and strategy II (right plot) for (a) ER networks with $\langle
  k \rangle=4$, $N=10^5$ and (b) SF networks with $\lambda=3.5$,
  $N=10^5$.  The arrows indicate the position of $p^*$ above which
  there is no phase transition. In the insets we show the ratio $R$
  between $\Delta S_1$ for $f=0.45$ and $\Delta S_1$ for $f=0.55$ for
  strategy I ($*$) and strategy II (x). All the simulation were done
  for $10^5$ networks realizations.\label{fig:f.delt}}
\end{figure}

When two groups compete, either group can use both random and targeted
strategies to influence the other group. Will the impact of these
strategies differ if the group using them is in the majority, as
opposed to being in the minority? Because the largest majority cluster
will have a larger average degree than the largest minority cluster,
we assume it will be harder to change the opinion of the majority than
the minority for $p<p_c$. In order to quantitatively
understand the effect of contrarians in both a minority group and a
majority group, we compute the relative change of the size of the
largest minority and majority clusters, $\Delta S_1$, given by
\[
\Delta S_1=(S_1^{\rm initial}-S_1^{\rm final})/S_1^{\rm initial},
\]
where $S_1^{\rm final}$ is the size of the largest $A$ cluster in our
final steady state and $S_1^{\rm initial}$ is the cluster before adding
the contrarians.  Notice that $f < 0.5$ ($f > 0.5$) means that the $A$
agents are minority (majority).  In Fig.~\ref{fig:f.delt} we plot
$\Delta S_1$ as a function of $p$ for $f=0.45$ (minority) and
$f=0.55$ (majority) under both strategies for both ER and SF networks.
From the plots we can see that below $p_c$ (marked by arrows in
the plots), $\Delta S_1$ is larger for minority than for majority for
the same value of $p$, under both strategies I and II, and for
the two networks used here, ER and SF. Thus, as argued above, the
minority groups are easier to convince than the majority
groups. Moreover, this phenomenon is more pronounced under the
targeted strategy than under the random. In the inset of
Fig.~\ref{fig:f.delt} we plot $R$, which is the ratio between $\Delta
S_1$ of minority and $\Delta S_1$ of majority, as a function of
$p$. As we can see for the ER network, the contrarians under a
targeted strategy are twice more effective when they are in the minority
group than when they are in the majority group, while under a random
strategy they are approximately 1.5 times more effective. We can see a
similar but more significant tendency in SF networks.  This agrees
with empirical fact, where majority groups always have more power than
minority groups, and thus it is easier for a majority group to change
the opinion of a minority group. We conclude that our model seems to mimic
well the two-group competition in the real world, and that it also
reveals some underlying complex phenomena associated with the
process.

\section{Conclusions}

In introducing contrarians into a system, we have used two strategies:
(i) random and (ii) targeted. Our contrarians hold a strong $B$
opinion and the system has two stable opinion $A$ clusters and opinion
$B$ clusters. We find that, for both strategies, the size of the
largest $A$ opinion cluster shrinks, as in a phase transition
phenomena. As the concentration of contrarians increases, the largest
$A$ cluster becomes smaller and smaller until it reaches zero at a
critical concentration $p_c$. Above $p_c$, the largest $A$ cluster
disappears. Our results show that the system undergoes a second order
phase transition for both control parameters $f$ and $p$, behavior
that resembles mean field percolation. We also find that, for both ER
and SF networks, the targeted strategy is more efficient than the
random strategy, because the targeted strategy always sends more
contrarians into the largest cluster than the random. Both strategies
effect more the minority group and much less the majority group. Note
that since SF networks have hubs, both strategies work better in SF
networks than in ER networks. Based on our results, we can answer the
questions we raise in the introduction. Free products and favors
(``bribes'') do effectively attract more supporters, but the most
effective strategy is to target those potential supporters with the
most connections and offer the free products and favors to them.

\section*{Acknowledgments}
We thank Jia Shao for discussions, and UNMdP, FONCYT-PICT 2008/0293,
ONR, DTRA, DFG, EU project Epiwork, and the Israel Science Foundation for financial support. 


\begin{thebibliography}{99}
\bibitem{note} In simulations of NCO model on 2D lattices, Shao et
  al\cite{NCO} suggested that it is in the same universality class as
  invasion percolation. We expect that our CO model on 2D lattices
  will be also in the same class as invasion percolation.
\end{thebibliography}

\newpage

\vspace{1cm}

\newpage

\newpage

\vspace{1cm}

\vspace{1cm}

\end{document}